\pdfoutput=1

\documentclass[aps,prl,twocolumn,superscriptaddress]{revtex4}
\usepackage{graphics}
\usepackage{epstopdf}
\usepackage{graphicx}
\usepackage{dcolumn}
\usepackage{bm}
\usepackage{amsmath}

\newcommand{\ceci}[1]{{}}

\newcommand{\be}{\begin{equation}}
\newcommand{\ee}{\end{equation}}
\newcommand{\beq}{\begin{eqnarray}}
\newcommand{\eeq}{\end{eqnarray}}

\def\nue{\mathrel{{\nu_e}}}
\def\numu{\mathrel{{\nu_\mu}}}
\def\nutau{\mathrel{{\nu_\tau}}}
\def\nux{\mathrel{{\nu_x}}}

\def\barnue{\mathrel{{\bar \nu}_e}}
\def\barnumu{\mathrel{{\bar \nu}_\mu}}
\def\barnutau{\mathrel{{\bar \nu}_\tau}}

\def \gta {\mathrel{\vcenter{\hbox{$>$}\nointerlineskip\hbox{$\sim$}}}}

\def\t13{\mathrel{{\theta_{13}}}}
\def\y12{\mathrel{{\tan^2 \theta_{12}}}}
\def\c2{\mathrel{{\chi^2 }}}

\def\msun{\mathrel{{M_\odot}}}

\newcommand{\n}{neutrino}
\newcommand{\ns}{neutrinos}

\newcommand{\sn}{supernova}
\newcommand{\sne}{supernovae}

\newcommand{\bh}{black hole-forming}
\newcommand{\nts}{neutron star-forming}

\newcommand{\f}{flux}

\newcommand{\sk}{SuperKamiokande}

\newcommand{\df}{DF}

\begin{document}


\title{Diffuse neutrino flux from failed supernovae}

\author{Cecilia Lunardini}
\affiliation{Arizona State University, Tempe, AZ 85287-1504}%
\affiliation{RIKEN BNL Research Center, Brookhaven National Laboratory, Upton, NY 11973}

 
\begin{abstract}
I study the diffuse flux of electron antineutrinos from stellar
collapses with direct black hole formation (failed supernovae).  This
flux is more energetic than that from successful supernovae, and
therefore it might contribute substantially to the total diffuse flux
above realistic detection thresholds.
%
The total flux might be considerably higher than previously thought, and approach the sensitivity of SuperKamiokande. For 
more conservative values of the parameters,  the flux from failed supernovae 
dominates for antineutrino energies above 30-45 MeV, with potential to give an
observable spectral distortion at Megaton detectors.
\end{abstract}                            
 
\maketitle

There is  confidence, in the neutrino astrophysics community, that
the diffuse flux (\df) of \ns\ from core collapse \sne\ will be
detected in the near future.  The current upper limit from the 50 kt water
tank of SuperKamiokande on diffuse electron antineutrinos,
$\phi_{\barnue} (E>19.3~{\rm MeV})< 1.4 - 2~{\rm cm^{-2} s^{-1}} $ at
90\% confidence level \cite{Malek:2002ns,Lunardini:2008xd}, 
already approaches theoretical predictions 
\cite{Lunardini:2005jf}. Within a decade or so, tens to hundreds of
events from the diffuse flux will be available from detectors of 0.1-1
Mt mass \cite{Barger:2007yw,Autiero:2007zj}, allowing steady progress
in the investigation of the physics of core collapse, of the
cosmological rate of \sne\ and of the properties of the \n.

So far, predictions for the \df\ have considered only the most common
 scenario: the collapse into a neutron star, with $ \sim 3 \cdot
 10^{53}~{\rm ergs}$ emitted in \ns\ of  average
 energy $E_{0} \sim 9 - 18$ MeV (see Eq. (\ref{dnde})).  Recently, detailed studies have appeared
 \cite{Liebendoerfer:2002xn,Sumiyoshi:2006id,Sumiyoshi:2007pp,Fischer:2008rh,Sumiyoshi:2008zw,Nakazato:2008vj}
 on the rarer case of \emph{ direct} collapse into a black hole
 without explosion, i.e., a failed \sn.  It was shown that the
 neutrino emission is somewhat more luminous and decidedly more
 energetic than for \nts\ collapse, due to the rapid contraction of the newly formed
protoneutron star preceding the black hole formation. Average energies are $E_0 \sim 20-24$ MeV for all \n\
 flavors.  This suggests that the hotter contribution of \bh\ collapses 
 to the \df\ might exceed that of \nts\ ones in part of
 the energy spectrum.

In this letter I study the diffuse flux from failed supernovae, with focus on its
the electron antineutrino
component, which is relevant for water Cherenkov detectors. The results
confirm the intuition that the flux from failed \sne\ might be
significant and bring the \df\ even closer to being finally detected.


Core collapse occurs for stars with mass $M\gta 8 M_{\odot}$
($M_{\odot}$ is the mass of the Sun), at an average rate of $R_{cc}(0)
\sim 10^{-4} ~{\rm Mpc^{-3} yr^{-1}}$ today \cite{Hopkins:2006bw} and
of
\be
R_{cc}(z)\simeq R_{cc}(0) (1+z)^\beta
\label{rcc}
\ee
(best fit $\beta =3.28$ \cite{Hopkins:2006bw}), up to redshift $z
\simeq 1$. The rate  flattens at larger $z$.
For $M = 8 - 25 M_{\odot}$ \cite{Woosley:2002zz} the collapse leads to an explosion -- the
observed explosion of $\sim$20$M_\odot$ star Sanduleak into SN1987A
\cite{SN87Adiscovery} supports this -- and to the formation of a
neutron star.

\begin{widetext}

\begin{figure}[htbp]
  \centering
 \includegraphics[width=0.45\textwidth]{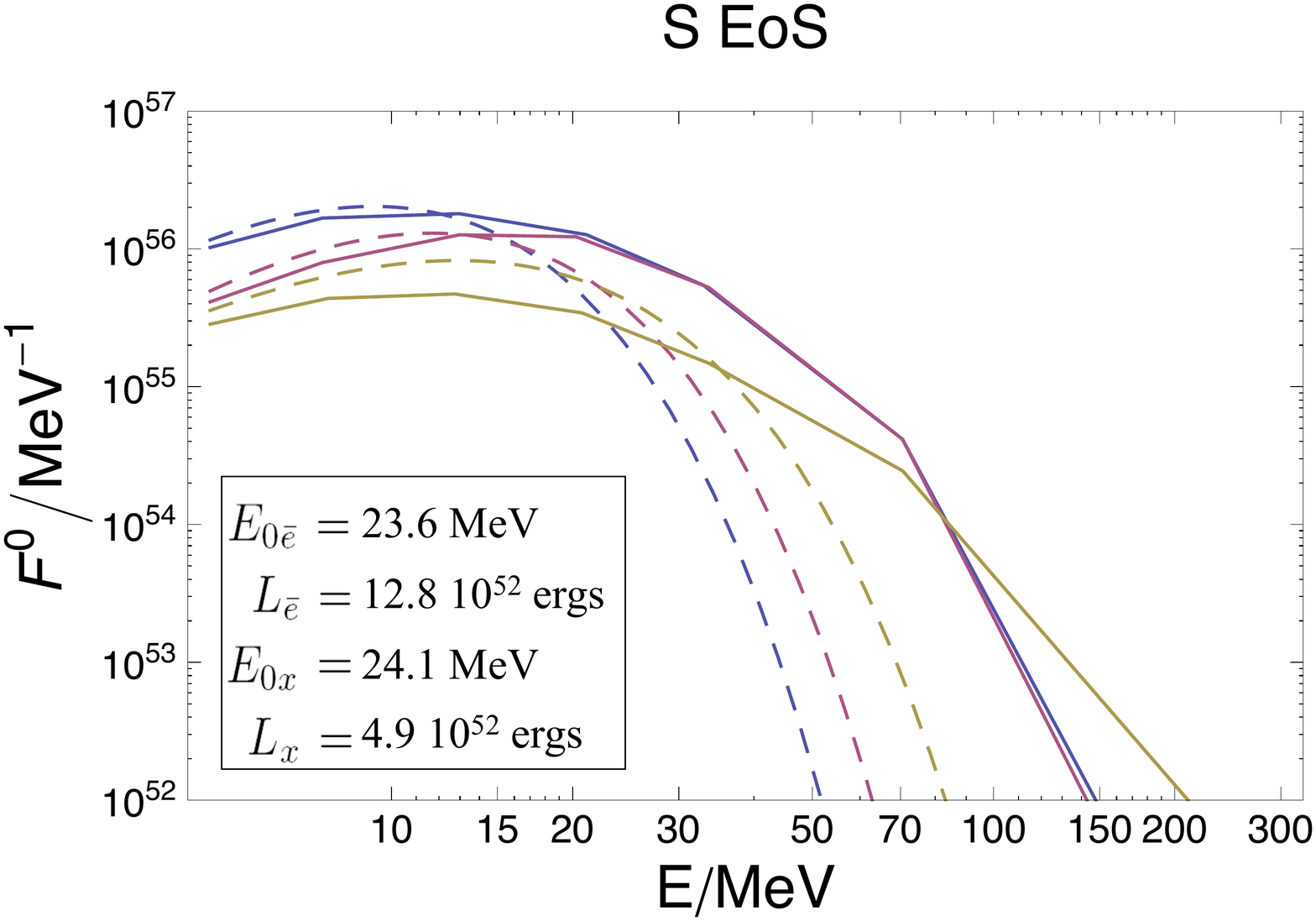}
 \includegraphics[width=0.45\textwidth]{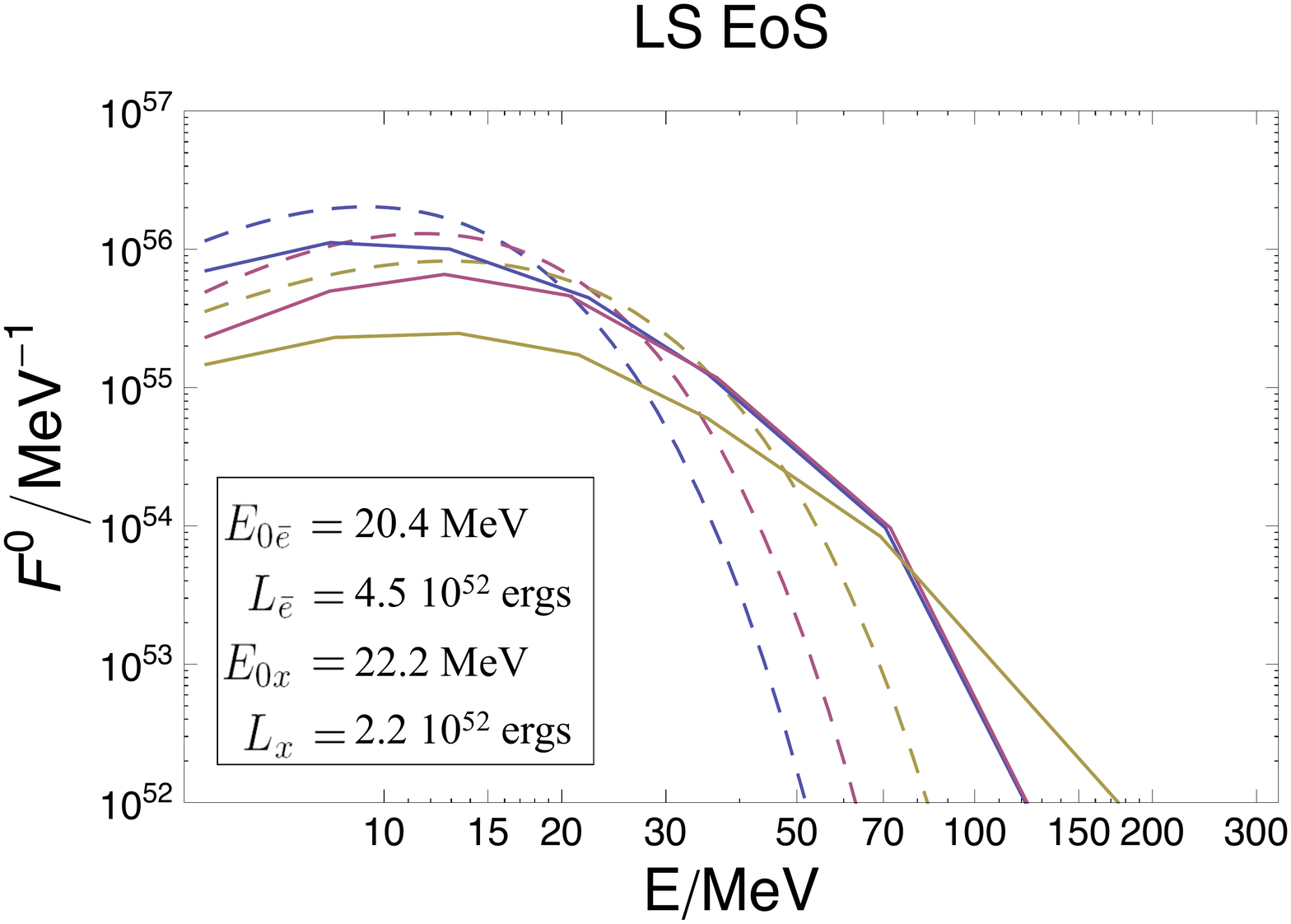}
   \caption{Neutrino fluxes at production inside the star for direct \bh\ collapse (solid, from \cite{Nakazato:2008vj}) ,
   and \nts\  collapse (dashed, Eq. (\ref{dnde})).  In both cases, the curves from upper to lower at 5 MeV correspond to $\nue$, $\barnue$, $\nux$.  
   For direct \bh\ collapse the \n\ spectra are shown for the Shen et al. (left panel) and Lattimer-Swesty (right) EoS.  For each, the \n\ luminosities and average energies are given (inserts).  See text for details.}
\label{spectra}
\end{figure}

\end{widetext}

Since stars are distributed in mass as $\phi(M) \propto
M^{-2.35}$ \cite{Salpeter:1955it}, one gets that \nts\ collapses are a
fraction $f_{NS}\simeq 0.78$ of the total.
Their \n\ output is roughly equipartitioned between the
 six 
 species: $\nue,\barnue,\numu,\barnumu,\nutau,\barnutau$ ($\numu,\barnumu,\nutau,\barnutau=\nux$ from here on).  At the production site, the flux in
 each species $w$, differential in energy, can be described as \cite{Keil:2002in}:
\be
F^0_w\simeq \frac{(1+\alpha_w)^{1+\alpha_w}L_w}
  {\Gamma (1+\alpha_w){E_{0w}}^2}
  \left(\frac{E}{{E_{0w}}}\right)^{\alpha_w}
  e^{-(1+\alpha_w)E/{E_{0w}}},
\label{dnde}
\ee
 where $\Gamma(x)$ stands for the Gamma function. Here $\alpha_w $ controls the spectral shape, $L_w $  is the time integrated luminosity and $E_{0 w}$ is
the average energy.  For illustration (fig. \ref{spectra}, dashed lines), here I use
the typical values \cite{Keil:2002in}: $E_{0 \bar e} = 15$ MeV, $E_{0
x} = 18$ MeV, $L_{\bar e}=L_x=5 \cdot 10^{52}$ ergs, $\alpha_{\bar
e}=3.5$ and $\alpha_x=2.5$.  After \n\ oscillations in the star (those
in the Earth are negligible \cite{Lunardini:2005jf}), the $\barnue$
flux is determined by the survival probability $\bar p$:
\beq
F_{\bar e} = \bar p F^0_{\bar e} + (1- \bar p ) F^0_{x}~, \hskip0.4truecm \bar p = 0-\cos^2 \theta_{12} 
\simeq 0 - 0.68 ~,
\label{mix}
\eeq
with $\theta_{12}\simeq 34^\circ$ \cite{Aharmim:2008kc}. The interval
 for $\bar p$ reflects the possible different oscillation scenarios,
 depending on the \n\ mass hierarchy, on the still unknown angle
 $\theta_{13}$ \cite{Dighe:1999bi} and on the effects of
 neutrino-neutrino scattering (see e.g., \cite{Dighe:2008dq,Chakraborty:2008zp}).  $\bar p$ can change during neutrino emission
 \cite{Schirato:2002tg}, but always remains in the interval
 (\ref{mix}), which therefore describes the time averaged probability.
 To illustrate how results can vary, I consider the extreme values of
 $\bar p$.  They can be taken as constant in energy, which is adequate
 \cite{Dighe:1999bi} for the experimentally relevant energy range
 ($E\gta 19.3$ MeV for SuperKamiokande \cite{Malek:2002ns}).

While \nts\ collapses (NSFCs)  have been widely studied, the evolution of higher
mass stars is more uncertain.  For $M \sim 25 - 40 \msun $ (13\% of
the total) a weaker explosion should occur, with a black hole formed
by fallback \cite{Woosley:2002zz}. Stars with $M \gta 40 \msun$ (a 9\%
fraction), would instead collapse into a black hole directly.
Simulations of such direct \bh\ collapses (DBHFCs)
\cite{Sumiyoshi:2006id,Sumiyoshi:2007pp,Fischer:2008rh,Sumiyoshi:2008zw}
show an emitted \n\ flux that is  more energetic and more luminous than
the NSFC case,  with especially high luminosity in
$\nue$ and $\barnue$ due to capture of
electrons and positrons on nucleons.
%
%

%
A ``stiffer'' equation of state (EoS) of nuclear matter
\cite{Sumiyoshi:2006id} and/or a smaller accretion rate of matter on the protoneutron star  \cite{Fischer:2008rh} correspond to more
luminous and hotter neutrinos. 
Here I use the fluxes from DBHFCs
as in fig. 5 of
ref. \cite{Nakazato:2008vj}. They are  are shown in fig. \ref{spectra} (solid lines and inserts). These fluxes were obtained for the $40 M_{\odot}$ progenitor in  \cite{woosley} with the stiffer  Shen et al. (S) EoS 
\cite{shenetal} (incompressibility $K=281$ MeV) and the softer  Lattimer-Swesty (LS) one  \cite{lattimer91generalized} (with $K=180$ MeV  \cite{Nakazato:2008vj}).  For the different progenitors considered in \cite{Nakazato:2008vj} 
results appear unchanged for the S EoS, while for the LS one 
the luminosity and average energy  may be lower by a factor of two and by 10-20\% respectively.
For the energy spectra, I use the same linear
interpolation of numerically calculated points as in \cite{Nakazato:2008vj}, 
which underestimates the \df\  in the \sk\ window by about 10-20\%, so my results 
are  conservative.
Ref. \cite{Nakazato:2008vj} shows that Eq. (\ref{mix}) applies to the DBHFC case, with the same extreme values of $\bar p$ being realized for the same oscillation parameters as in the NSFC case.


 I calculated the neutrino fluxes from NSFCs and DBHFCs, and the total \df\ for a schematic
two-population scenario, with a fraction $f_{NS}$ ($f_{BH}=1-f_{NS}$)
of identical \n\ emitters of the NSFC (DBHFC) type \footnote{the \n\
emission in the case of black hole formation by fallback -- for which
detailed studies lack -- should be qualitatively similar to the \nts\
case because the timescale of the black hole formation is expected to
exceed that of \n\ emission \cite{Woosley:2002zz}.}.  Generalizing the
single population formula (e.g., \cite{Ando:2004hc}) one gets the
total diffuse $\barnue$ flux at Earth, differential in energy and area:
\begin{widetext}
\be
\Phi(E)=\frac{c}{H_0}\int_0^{z_{ max}} R_{cc}(z)\left[ f_{NS} F^{NS}_{\bar e}(E(1+z)) + (1-f_{NS}) F^{BH}_{\bar e}(E(1+z))\right]\frac{{d}
z}{\sqrt{\Omega_{ m}(1+z)^3+\Omega_\Lambda}}~,
\label{difflux}
\ee
\end{widetext}
where $\Omega_{ m}=0.3$ and $\Omega_\Lambda=0.7$ are the fractions of
the cosmic energy density in matter and dark energy; $c$ is the speed
of light and $H_0$ is the Hubble constant. I took the parameters
$R_{cc}(0) = 10^{-4} ~{\rm Mpc^{-3} yr^{-1}}$, $\beta=3.28$
and $z_{max}=4.5$ \cite{Hopkins:2006bw} (results depend weakly on $z_{max}$, at the level of $\sim 7\%$ or less for $z_{max}\gta 3$
\cite{Ando:2004hc}).  To parametrize the uncertainty in $f_{NS}$ I
take the interval $f_{NS}= 0.78 - 0.91$,  corresponding to a mass
$25-40 \msun$ as upper limit for neutron star-forming collapse.

\begin{widetext}

\begin{figure}[htbp]
  \centering
 \includegraphics[width=0.39\textwidth]{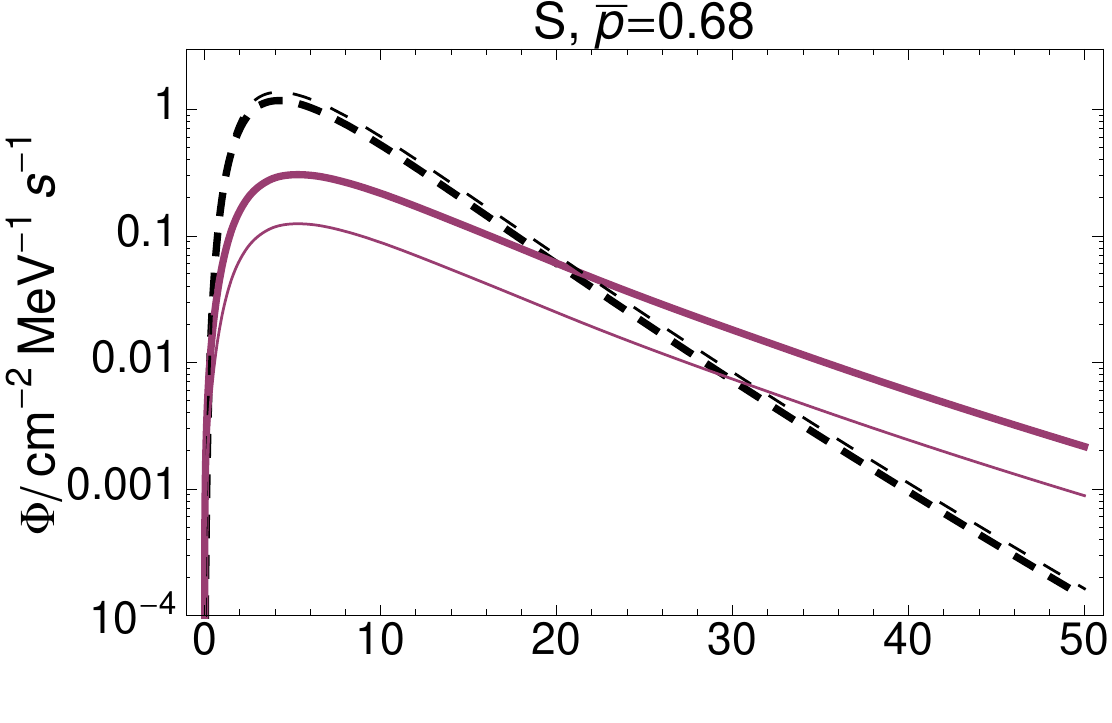}
 \includegraphics[width=0.39\textwidth]{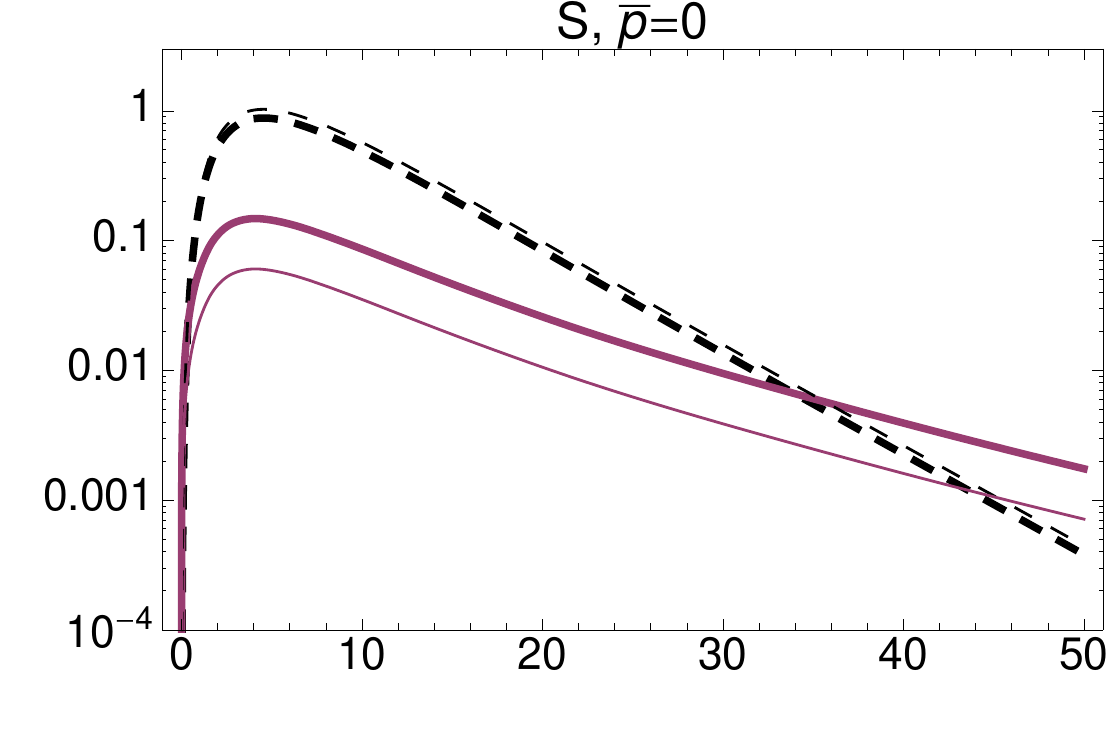}
 \includegraphics[width=0.39\textwidth]{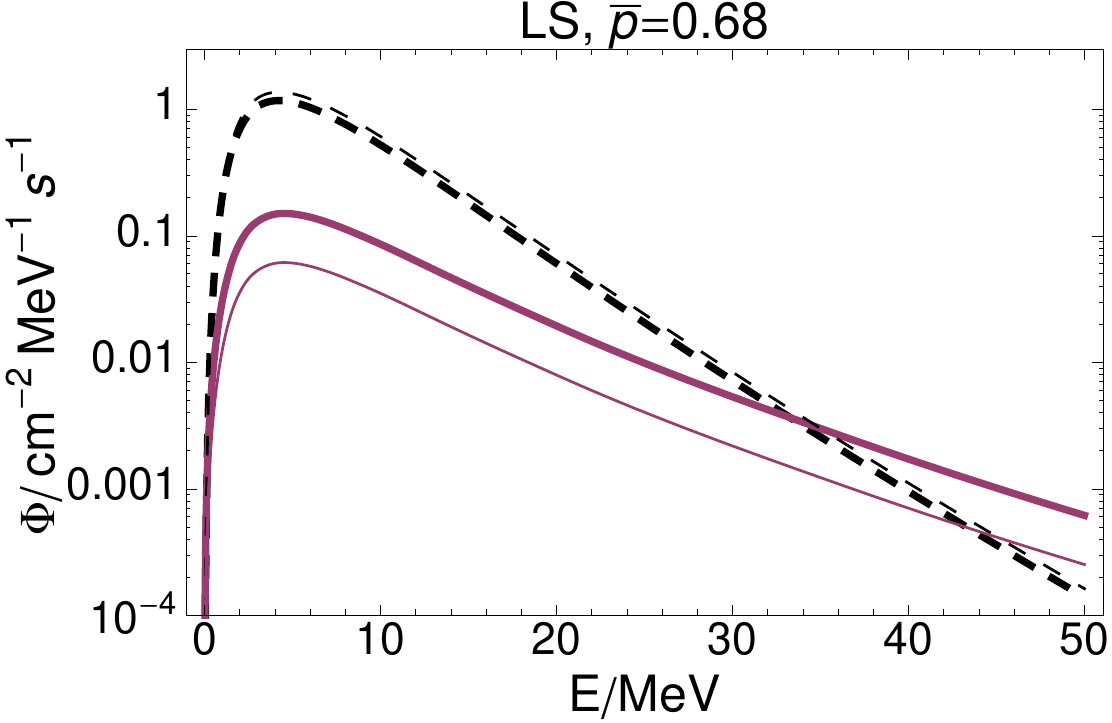}
 \includegraphics[width=0.39\textwidth]{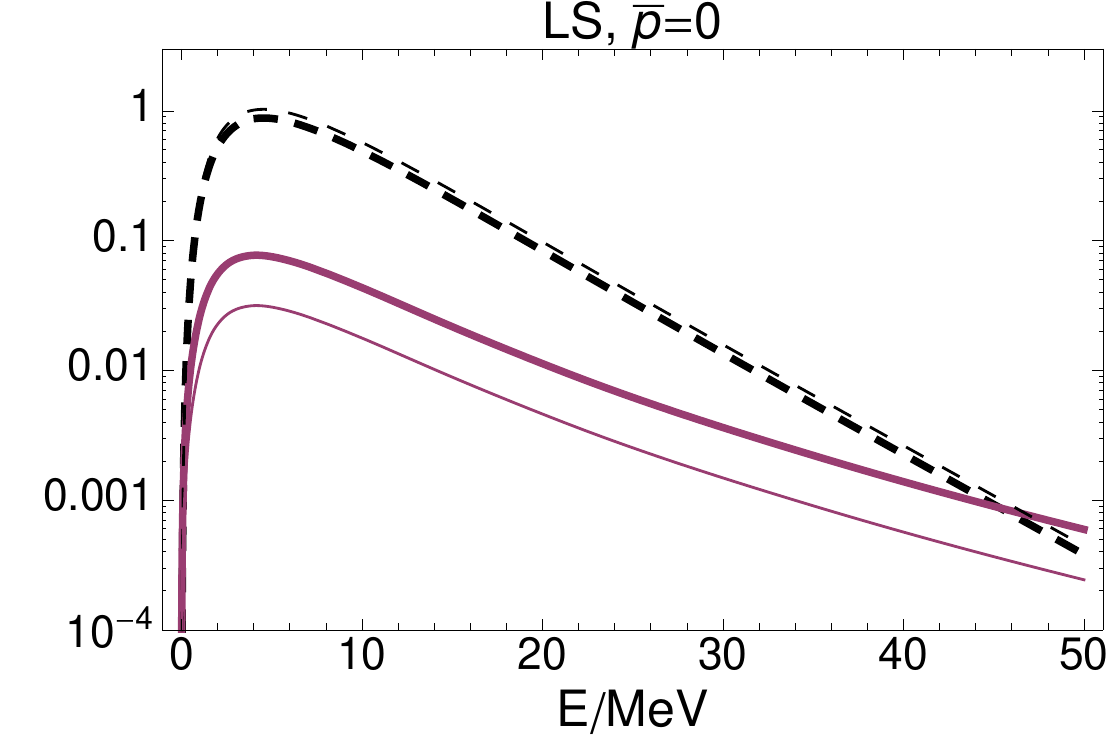}
   \caption{The diffuse \f\ of $\barnue$ at Earth from \nts\
   collapses (dashed curves) and direct \bh\ ones (solid), for different
   EoS and survival probability $\bar p$.  Direct black hole -forming collapses
   are assumed to be 22\% (thick curves) or 9\% (thin curves) of the total (see Eq. (\ref{difflux})). }
\label{spectradiff}
\end{figure}

\end{widetext}

Results are shown in fig. \ref{spectradiff}.  The diffuse flux from  NSFCs, $\Phi_{NS}$, 
is maximum at $E \sim 5-6$
MeV, with an exponential decay at higher energy
\cite{Lunardini:2006pd}.  The  contribution from DBHFCs, $\Phi_{BH}$, has hotter
spectrum, and thus is increasingly important at higher
energy. \ceci{somewhere I should say that high redshift contributions
may be interesting.... }  Oppositely to  $\Phi_{NS}$,
$\Phi_{BH}$ is larger for  minimal permutation ($\bar p=0.68$)
\cite{Nakazato:2008vj}, because of the high original $\barnue$ flux.
The dependence of the original fluxes on the EoS is evident in $\Phi_{BH}$.

\begin{widetext}

\begin{figure}[htbp]
  \centering
 \includegraphics[width=0.45\textwidth]{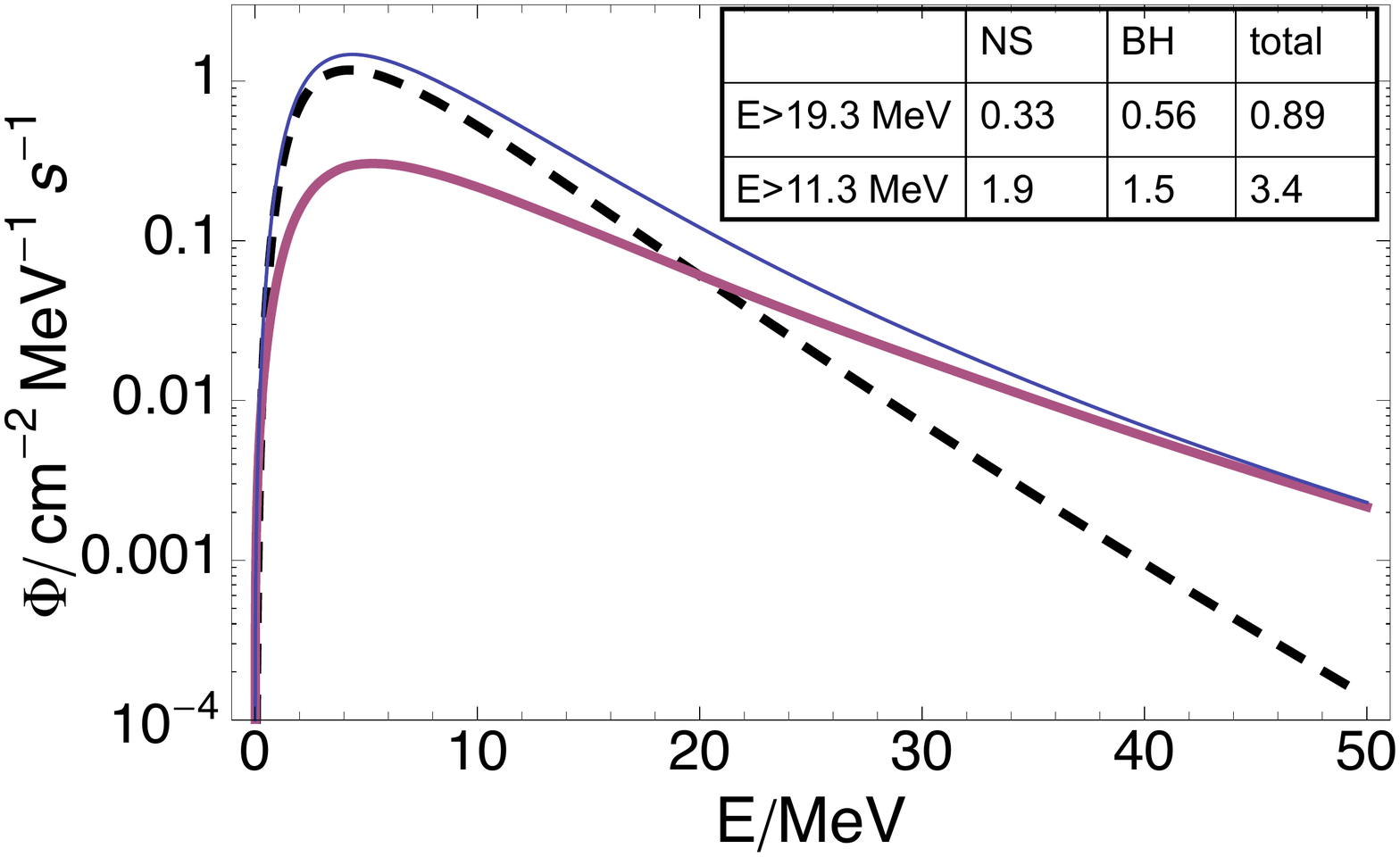}
\includegraphics[width=0.45\textwidth]{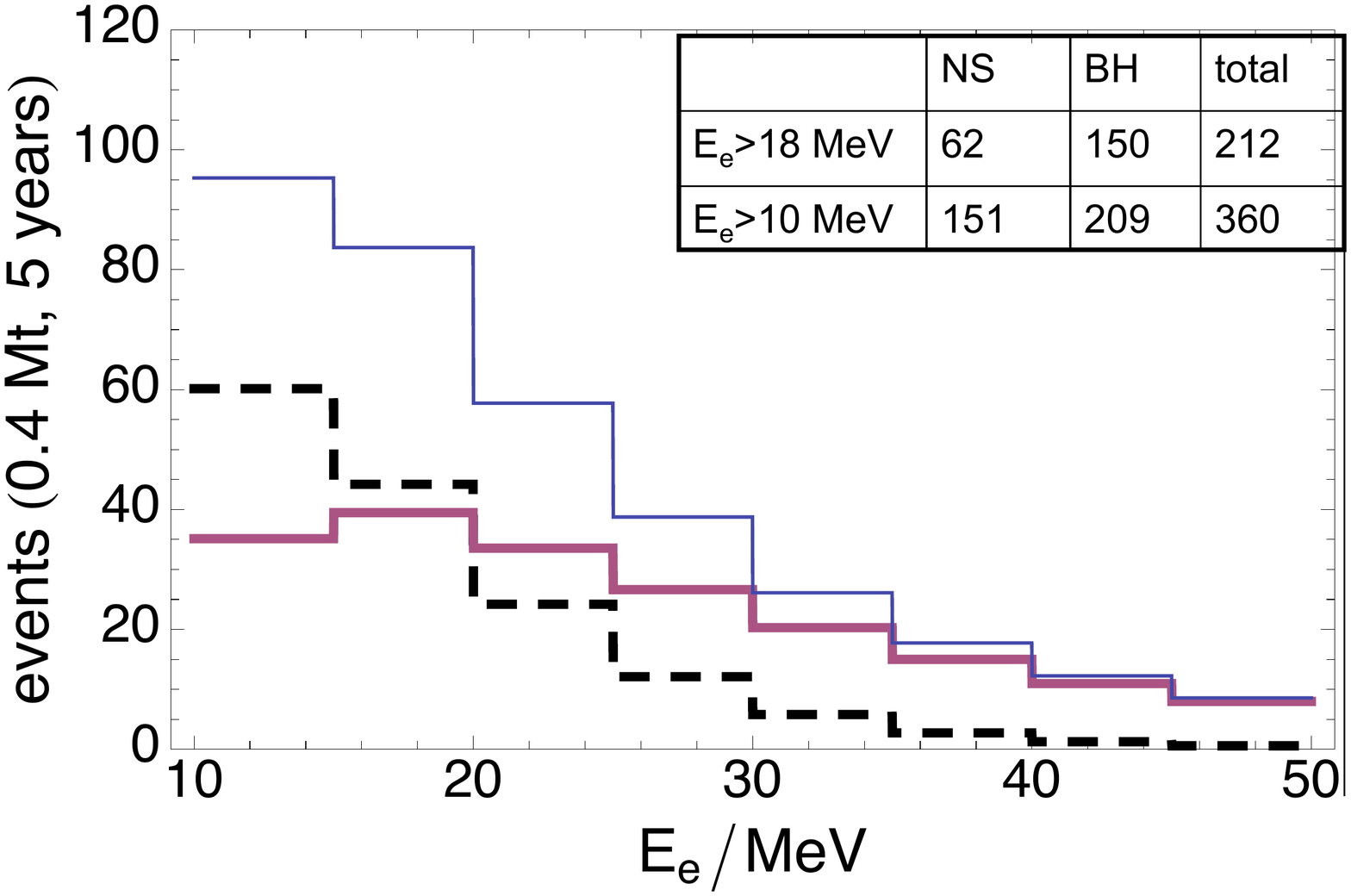}
   \caption{Left:  the largest possible $\barnue$ diffuse flux  from direct \bh\ collapses  (solid, thick line), compared to that from \nts\ ones (dashed), from  fig. \ref{spectradiff}. The total flux is shown too (thin).  Integrated fluxes above two thresholds of interest are given, in ${\rm cm^{-2} s^{-1}}$.   Right: same figure for inverse beta decay events in water with 2 Mt$\cdot$yr exposure.
   $E_e$ is the positron energy.
   I used  $R_{cc}(0)=10^{-4}~{\rm Mpc^{-3} yr^{-1}}$ for the core collapse rate today and ref. \cite{Strumia:2003zx} for the detection cross section. }
\label{summary}
\end{figure}
\end{widetext}

Fig. \ref{spectradiff} shows that $\Phi_{BH}$ might
dominate already at $E \sim 22$ MeV, implying a strong effect at
SuperKamiokande. For the most favorable parameters 
the total flux 
above 19.3 MeV more than doubles compared to
100\% NSFCs, (fig. \ref{summary}), reaching a value
($\Phi\simeq 0.89 ~{\rm cm^{-2} s^{-1}}$) tantalizingly close to the
current upper limit. The enhancement of the event rate is even larger,
thanks to the $\sim E^2$ dependence of the detection cross section.  Thus,
the \df\ might be closer to detection than previously thought, within
the reach of improved searches at \sk.

It is more likely, however, that $\Phi_{BH}$ exceeds $\Phi_{NS}$ only
above 30-40 MeV (fig. \ref{spectradiff}).  Its effect would be below
the sensitivity of \sk\ -- which would therefore place limits on \ns\ from DBHFCs
 -- but might be visible with the 1 Mt planned Cherenkov
detectors, where ${\mathcal O}(10)$ events are expected in this energy
interval for a few years running time.  Besides the excess in event
rate, which suffers normalization uncertainties, the DBHFC diffuse flux could be
visible for the spectral distortion that it produces.  The lower
threshold ($\sim 11$ MeV) of a liquid scintillator
\cite{MarrodanUndagoitia:2006re} or Gadolinium-loaded water detector
\cite{Beacom:2003nk} could allow to see a break in the energy spectrum at $\sim
$20 MeV, that might escape a pure water detector.

If all parameters conspire to maximally suppress it,  $\Phi_{BH}$ might be invisible even for a Mt detector at least in the first
few years running time.  A negative result would then
constrain the parameter space strongly.


To illustrate how results change for a more energetic NSFC flux I have repeated the calculations with $E_{0x}=22$ MeV; in what follows I discuss how results compare to those in fig.  \ref{spectradiff}. For $\bar p =0.68$ differences are only minimal relative to fig.  \ref{spectradiff}. 
This is because the $\barnue$ flux at Earth is dominated
by the original $\barnue$ flux, which for the DBHFCs  is markedly
more energetic than in the NSFC case.  For the S EoS $\Phi_{BH}$ exceeds $\Phi_{NS}$ above 22 MeV (38 MeV) for $f_{NS}=0.78$
($f_{NS}=0.91$).  Instead, for the LS EoS the  two components are
comparable at energy of 44 MeV or higher. This slight worsening
compared to fig. \ref{spectradiff} may make the difference between a
positive or negative signal of $\Phi_{BH}$ at an experiment.
 For
$\bar p=0$ the distinctive original $\barnue$ flux from DBHFC 
does not contribute to the $\barnue$ flux at Earth, and so the main
signature of direct black hole formation is lost. One may see a slight excess
flux at $E \sim 45-50$ MeV only for the S EoS and
$f_{NS}=0.78$.

If the flux excess due to $\Phi_{BH}$ appears only above 30 MeV, it might
partially be masked by the invisible muon and atmospheric \n\
backgrounds, which are strong at that energy \cite{Malek:2002ns}. I modeled these following \cite{Fogli:2004ff} and
taking a 100\% flux excess in the 30-35 MeV bin, relative to a
theoretical model or fit to the data with $f_{NS}=1$.  For pure water,
where invisible muons dominate, the excess would not be
statistically significant in the single bin, but might be distinguishable 
%
with a  fit of the spectrum of events.
With the reduced background allowed by Gadolinium,
$3\sigma$ significance in the single bin would be achieved with about
12 Mt$\cdot$yr exposure, with a lower exposure needed
if a spectral fit is done.


To summarize, 
the diffuse flux of \ns\ from failed supernovae may be significant, 
at a level detectable at \sk\ or at Mt scale detectors.
While  conclusions are limited by 
uncertainties, 
it is hoped that 
this letter will serve as a motivation for the development of more realistic predictions of the diffuse flux, which would be of great service to the experimental community.

I thank H.T.~Janka,  A. Mezzacappa and O.~L.~G.~Peres for useful discussions, and ASU and RBRC for support.

\end{document}